\documentclass[sigconf]{acmart}

\AtBeginDocument{%
  }

\setcopyright{none}
\usepackage{tcolorbox}

\usepackage{hyperref}
\usepackage{xcolor}
\usepackage{booktabs}
\usepackage{multirow}   
\usepackage[table]{xcolor} 
\usepackage{graphicx}  
\usepackage{wrapfig}
\usepackage{amsmath} 
\usepackage[utf8]{inputenc}
\usepackage[english]{babel}
\tcbuselibrary{breakable}

\usepackage[table,dvipsnames]{xcolor} 

\usepackage{tcolorbox}
\usepackage{algorithm}
\usepackage{enumitem}
\tcbuselibrary{skins}
\usepackage{algpseudocode}
\usepackage{geometry}
\usepackage{xcolor}
\usepackage{tabularx}
\usepackage{subcaption}
\usepackage{cleveref}
\usepackage{url}
\usepackage{footmisc}

\newcommand{\poschange}[1]{\textcolor{red}{\textbf{+}#1}}
\newcommand{\negchange}[1]{\textcolor{green!60!black}{\textbf{–}#1}}

\definecolor{problembg}{HTML}{E0EFFF}  
\definecolor{goodalgobg}{HTML}{E0F8E0} 
\definecolor{badalgobg}{HTML}{FFF0E0}  
\definecolor{analysisbg}{HTML}{F5F5F5} 

\newtcolorbox{paperbox}[2][]{
    colback = #2,
    colframe = #2!50!black,
    fonttitle = \bfseries,
    title = #1,
    arc = 2mm,
    boxrule = 1pt,
    left = 5mm,
    right = 5mm,
    top = 3mm,
    bottom = 3mm,
    breakable 
}

\newboolean{showcomments}
\setboolean{showcomments}{true}
\ifthenelse{\boolean{showcomments}}
 { \newcommand{\mynote}[2]{
      \fbox{\bfseries\sffamily\scriptsize#1}
        {\small$\blacktriangleright$\textsf{\emph{#2}}$\blacktriangleleft$}}}
        { \newcommand{\mynote}[2]{}}

\newcommand{\tr}{\operatorname{tr}}
\newcommand{\E}{\mathbb{E}}
\newcommand{\Cov}{\operatorname{Cov}}
\newcommand{\Var}{\operatorname{Var}}

\begin{document}

\title{Dynamic Stability of LLM-Generated Code}


\author{Prateek Rajput}
\email{prateek.rajput@uni.lu}
\orcid{0000-0001-2345-6789}
\affiliation{%
  \institution{University of Luxembourg}
  \country{Luxembourg}
}

\author{Abdoul Aziz Bonkoungou}
\email{abdoul.bonkoungou@uni.lu}
\orcid{0009-0002-2361-485X}
\affiliation{%
  \institution{University of Luxembourg}
  \country{Luxembourg}
}

\author{Yewei Song}
\email{yewei.song@uni.lu}
\orcid{0000-0002-6314-7515}
\affiliation{%
  \institution{University of Luxembourg}
  \country{Luxembourg}
}

\author{Abdoul Kader Kabore}
\email{abdoulkader.kabore@uni.lu}
\orcid{0000-0002-3151-9433}
\affiliation{%
  \institution{University of Luxembourg}
  \country{Luxembourg}
}

\author{Iyiola E. Olatunji}
\email{emmanuel.olatunji@uni.lu}
\orcid{0000-0002-0391-9202}
\affiliation{%
  \institution{University of Luxembourg}
  \country{Luxembourg}
}

\author{Jacques Klein}
\email{jacques.klein@uni.lu}
\affiliation{%
  \institution{University of Luxembourg}
  \country{Luxembourg}
}

\author{Tegewende Bissyande}
\email{tegewende.bissyande@uni.lu}
\affiliation{%
  \institution{University of Luxembourg}
  \country{Luxembourg}
}

\renewcommand{\shortauthors}{Trovato et al.}

\begin{abstract}
Current evaluations of LLMs for code generation emphasize \emph{functional correctness}, overlooking the fact that functionally correct solutions can differ significantly in algorithmic complexity. For instance, an $(O(n^2))$ versus $(O(n \log n))$ sorting algorithm may yield similar output but incur vastly different performance costs in production. This discrepancy reveals a critical limitation in current evaluation methods: they fail to capture the behavioral and performance diversity among correct solutions.

To address this, we introduce a principled framework for evaluating the dynamic stability of generated code. We propose two metrics derived from opcode distributions: \textbf{Static Canonical Trace Divergence (SCTD)}, which captures algorithmic structure diversity across generated solutions, and \textbf{Dynamic Canonical Trace Divergence (DCTD)}, which quantifies runtime behavioral variance. Their ratio, the \textbf{Behavioral Expression Factor (BEF)}, serves as a diagnostic signal: it indicates critical runtime instability when BEF $\ll$ 1 and functional redundancy when BEF $\gg$ 1.

Empirical results on BigO(Bench) and CodeContests show that state-of-the-art LLMs exhibit significant algorithmic variance even among functionally correct outputs. Notably, increasing sampling temperature improves pass@1 rates but degrades stability, revealing an unrecognized trade-off: searching for correct solutions in diverse output spaces introduces a "penalty of instability." 
Our findings call for stability-aware objectives in code generation and new benchmarks with asymptotic test cases for robust, real-world LLM evaluation. Our anonymized artifacts is available\footnotemark\label{artifactsrepo}
\footnotetext{\url{https://github.com/azizYaaba/LLMCodeStability}}.


\end{abstract}



\maketitle

\section{Introduction}

The advent of Large Language Models (LLMs) is rapidly reshaping the software development landscape, enabling automated code generation at unprecedented scale and speed~\cite{zhang2023survey, chen2021evaluating, li2022competition}. 
As these systems transition from research prototypes to production tools, their evaluation metrics must evolve. While the current dominant paradigm emphasizes functional correctness, typically measured using the $pass@k$ metric~\cite{xu2022systematic}, or syntactic/semantic similarity \cite{chen2021evaluating, xu2022systematic, ren2020codebleu, mastropaolo2021studying}, this over-reliance is increasingly problematic. For instance, the $pass@k$ metric suffers from known statistical limitations and offers no insight into the efficiency, maintainability, or robustness of the generated code~\cite{yeo2024framework}. As a result, this narrow evaluation overlooks a critical dimension of code quality: \textbf{runtime stability}. Recent evolutionary AI systems for code generation, while enhancing performance, further compound output variability, underscoring the urgency for robust stability assessment~\cite{pordanesh2025robustness}.

Runtime stability is fundamental to predictable and efficient software execution. Unstable algorithms can introduce performance bottlenecks, trigger denial-of-service vulnerabilities through resource exhaustion~\cite{crosby2003denial}, and violate timing constraints which have severe consequences in safety-critical systems such as avionics and automotive control~\cite{buttazzo1997hard}. Beyond performance, instability inflates operational costs by hindering efficient resource provisioning, increasing energy consumption~\cite{tiwari2002power}, and masking performance bugs that only emerge under specific conditions~\cite{ball1994optimally}. Previous research has extensively explored runtime stability in traditional, human-written software, focusing on performance profiling, algorithmic complexity, and identifying bottlenecks \cite{ball1994profiling, cormen2022introduction, knuth1997art}. 
These efforts have yielded invaluable techniques for optimizing code and ensuring predictable behavior. 
However, this established body of knowledge largely predates the era of automated code generation.

\begin{figure*}[ht]
\centering
\begin{subfigure}[t]{0.42\linewidth}
    \resizebox{!}{10cm}{\input{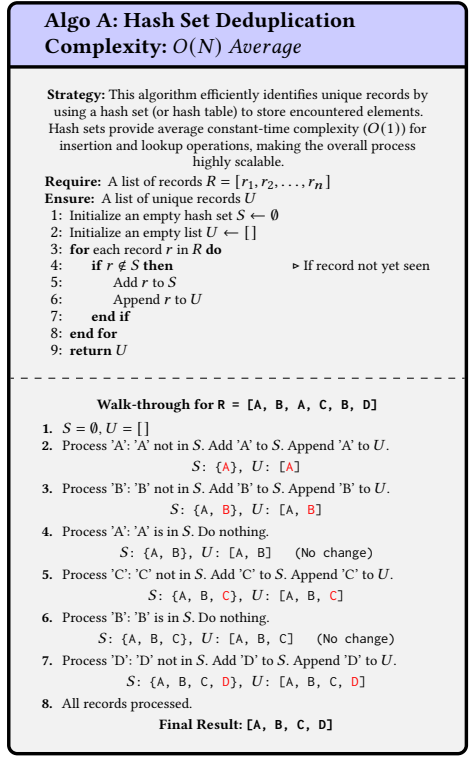}}
    \caption{Algorithm A}
\end{subfigure}
\hfill
\begin{subfigure}[t]{0.50\linewidth}
    \resizebox{!}{10cm}{\input{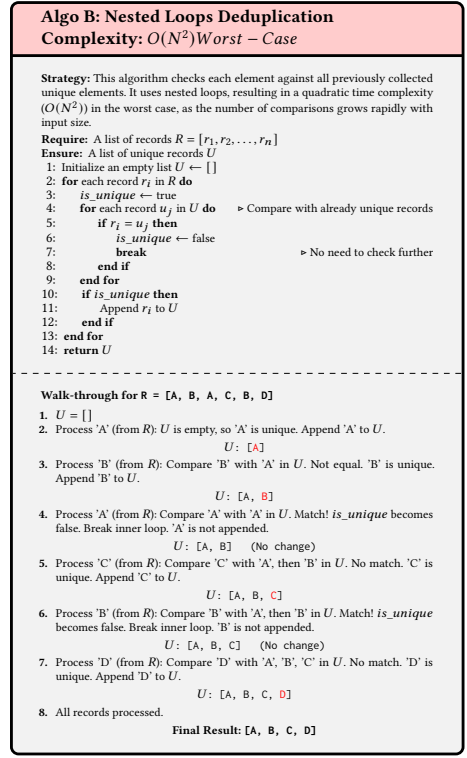}}
    \caption{Algorithm B}
\end{subfigure}
\caption{Two functionally correct algorithms for deduplication, but with vastly different runtime characteristics. Algorithm A scales efficiently, while Algorithm B suffers from quadratic-time degradation.}
\label{fig:motivating-algos}
\end{figure*}

Even small variations in LLM-generated logic can lead to dramatically different execution behaviors, ranging from subtle slowdowns to catastrophic failures under load. 
As LLMs become deeply integrated into professional development workflows, the absence of metrics that capture runtime behavior becomes a critical liability. While functional correctness may suffice for prototyping, it fails to expose the broader operational risks. 
Uncontrolled variation in algorithmic choices, edge-case performance, or memory usage across seemingly equivalent generations can result in outages, cost overruns, or security vulnerabilities in production. This positions the measuring of dynamic stability of LLM-generated code as an operational necessity for building reliable, scalable, and secure AI-driven software systems.

\noindent
{\bf Motivating Example. }
Consider, for instance, a common task like deduplicating a large dataset of customer records as shown in \Cref{fig:motivating-algos}. 
An LLM might be prompted to provide an algorithm for this.
It could generate Algorithm A, which uses a hash set for efficient, near constant-time lookups on average, offering robust performance even as data volume scales.
Alternatively, the same LLM might propose Algorithm B, which employs nested loops with a linear search for each element, leading to a quadratic time complexity. 
While both algorithms are functionally "correct", Algorithm B's performance would degrade exponentially with increasing input size, consuming excessive CPU cycles and memory.
In a production environment, deploying Algorithm B even if it passes unit tests on limited samples could lead to application freezes, timeouts, or a significant spike in cloud computing costs, illustrating a direct and quantifiable consequence of runtime instability.

\noindent
{\bf This paper. }
Our study addresses a critical gap in the evaluation of code-generating LLMs by introducing a novel framework for assessing their dynamic stability.

Understanding the true algorithmic structure and runtime behavior of code, especially that generated by LLMs, requires looking beyond surface-level syntax or abstract syntax trees. 
These high-level representations often fail to capture the subtle, yet critical, differences in control flow, memory access patterns, or fundamental operations that dictate a program's performance and stability. 
We propose to rely on opcodes: a low-level, platform-independent intermediate representation of code~\cite{aho2007compilers}. 
By analyzing opcode distributions, we can gain a granular understanding of how an algorithm truly executes, reflecting its intrinsic efficiency and resource utilization, rather than just its syntactic form~\cite{muchnick1997advanced}.

We propose two principled, bounded metrics derived directly from these opcode distributions: Static Canonical Trace Divergence (SCTD), which quantifies the diversity of algorithmic structures across multiple generated solutions; and Dynamic Canonical Trace Divergence (DCTD), which measures the runtime behavioral variance of these solutions across a test suite. 
Furthermore, our Behavioral Expression Factor (BEF), calculated as the ratio of DCTD to SCTD, uniquely captures critical cases where seemingly minor structural differences, as seen in opcodes, lead to disproportionately large behavioral divergences. This enables us to identify problematic instabilities often overlooked by traditional correctness-focused metrics.


\noindent
{\bf Contributions.} Our main contributions are as follows.
\begin{enumerate}[leftmargin=*]
    \item We introduce a novel framework to quantify code stability using two principled metrics: \textbf{SCTD} (structural) and \textbf{DCTD} (behavioral), derived from canonical opcode traces. We further define the \textbf{BEF}, a diagnostic ratio (\texttt{SCTD/DCTD}) that reveals both functional redundancy and runtime instability.
    
    \item Through a large-scale empirical study, we identify and name the "\textit{penalty of instability}": a trade-off where increasing temperature to diversify solution search space improves \textit{pass@k} rates but systematically degrades runtime stability, exposing a hidden cost of temperature-based sampling.

    \item We demonstrate that correctness-only evaluations and simple prompt controls are insufficient to manage this instability, advocating for dynamic stability as a first-class objective in model training and benchmarking, especially with asymptotic test cases.
\end{enumerate}

\section{Methodology}
Our empirical protocol is designed to systematically quantify the algorithmic and dynamic stability of code artifacts generated by LLMs. The fundamental task under investigation is thus \textbf{code generation}. For each experimental run, the \textit{input} is a prompt containing a complete problem description sourced from our benchmarks, including its formal statement, input/output specifications, and constraints. The expected \textit{output} is a compilable Python source code function that is intended to solve the given problem. For each problem-model pair, we generate a corpus of candidate solutions to analyze their collective stability.

A critical prerequisite for our stability analysis is the establishment of functional correctness. We define a single generated artifact as \textit{functionally correct} if it successfully passes all public unit tests provided by the source benchmark. Our stability analysis is therefore \textit{conditioned} on this correctness (although we also measure it for cases where not all solutions are correct); we focus primarily on the cohort of solutions that satisfy this criterion, enabling us to isolate the variance in algorithmic strategy and runtime behavior from rudimentary functional failures. This allows us to investigate a more nuanced question: \textit{among a set of demonstrably correct solutions, how stable are their underlying computational properties?} The following sections detail the specific datasets, models, and our novel metrics (SCTD and DCTD) that constitute this evaluation framework.

\subsection{Overview}

\begin{figure*}[ht]
    \centering
    \includegraphics[width=0.7\textwidth]{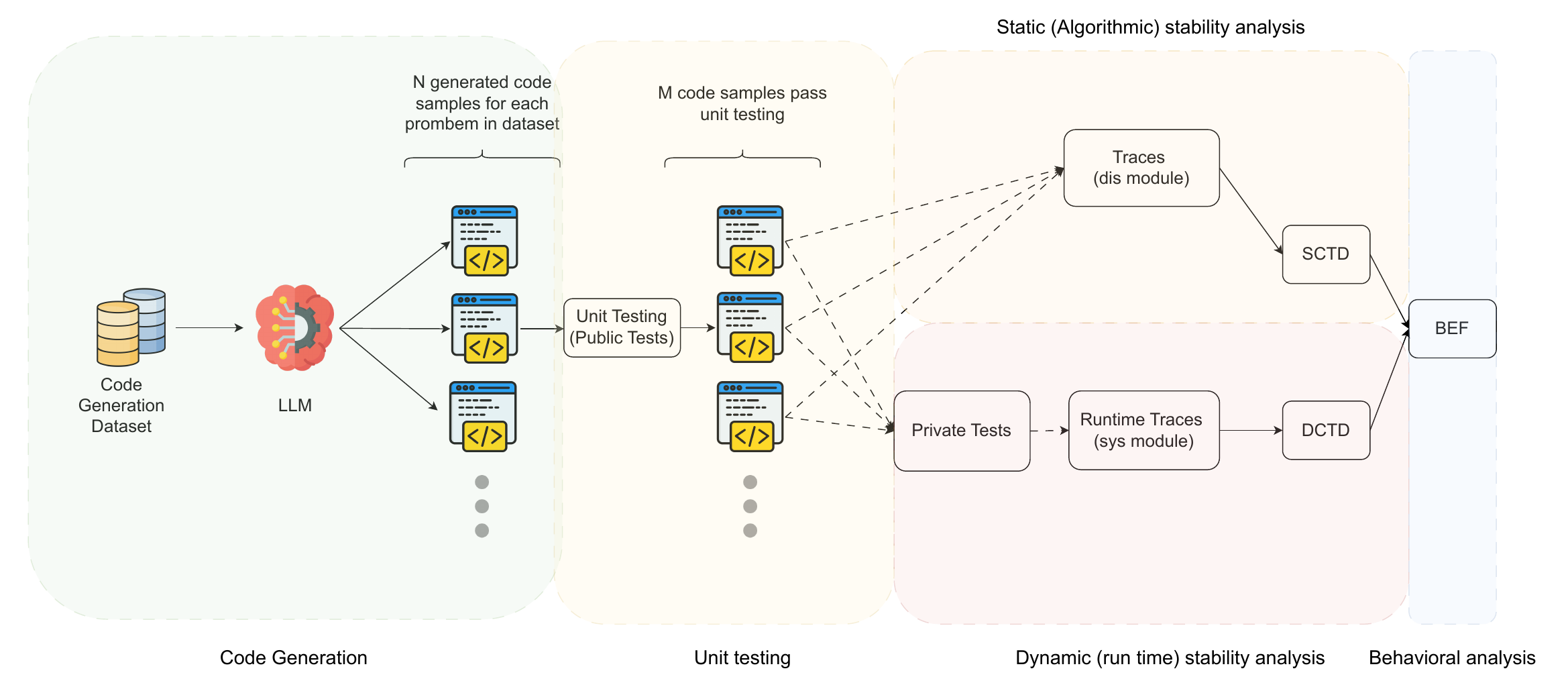}
    \caption{Pipeline illustrating the end-to-end process for computing SCTD, DCTD, and BEF. Static stability is assessed through analysis of Python bytecode using the \texttt{dis} module (SCTD), while dynamic stability is evaluated through runtime traces (DCTD).}
    \label{fig:pipeline_ini}
    \vspace{-1em}
\end{figure*}
\Cref{fig:pipeline_ini} shows a schematic of our end-to-end evaluation pipeline, which proceeds as follows. First, for a given problem, we prompt an LLM to generate a set of $n$ candidate solutions. These solutions are then systematically validated against the benchmark's public unit tests to establish a baseline of functional correctness. Based on the number of passing solutions, $m$, we classify each generation set into one of three distinct cohorts: \texttt{all\_success} (where $m=n$), \texttt{some\_success} ($0 < m < n$), or \texttt{all\_fail} ($m=0$). Although our metrics are computed for all cases, this paper's primary analysis concentrates on the \texttt{all\_success} cohort.

For the $m$ functionally correct solutions within a set, we then compute our stability metrics. To derive the SCTD, we perform a static analysis by compiling each solution into Python bytecode via the \texttt{dis} module, yielding a static opcode frequency distribution. To derive the DCTD, we conduct a dynamic analysis by executing each correct solution against a suite of $r$ private test cases. We instrument the runtime using the \texttt{sys} module's tracing capabilities to capture a dynamic opcode frequency distribution for each test execution. These frequency counts are then normalized into Probability Mass Functions (PMFs) over the $d$-dimensional opcode vocabulary. To better capture computational expense, we also create a parallel set of cost-weighted PMFs by applying a heuristic weight $w_i$ to each opcode, as detailed below. Both the unweighted and weighted PMFs form the basis from which \textbf{SCTD} and \textbf{DCTD} are respectively calculated.

\subsubsection{Opcode Distributions as a Proxy for Algorithmic Behavior}

The core of our methodology is the use of opcode distributions, derived from compiled bytecode and execution traces, as a proxy for the underlying algorithmic strategy. We operationalize this by constructing PMFs over the opcodes. To account for varying computational costs, each opcode is assigned a weight representative of its runtime complexity. We selected heuristic weights of 1, 10, and 100 to differentiate these costs; for instance, simple stack operations like POP\_TOP are weighted at 1, operations with linear complexity such as BUILD\_LIST are weighted at 10, and computationally intensive opcodes like BINARY\_MATRIX\_MULTIPLY are assigned a weight of 100. This approach enables a principled comparison of algorithmic strategies by measuring the statistical distance between the resulting weighted PMFs.

\subsubsection{Canonical Representation of Code Behavior}

We define key data tensors to represent the opcode profile of each solution, where: Let $m$ be the number of generated solutions that passed the unit tests; let $d$ be the number of distinct opcodes; let $r$ be the number of test cases; and let $w_i \in {1, 10, 100}$ be a tunable cost weight assigned to opcode $i$. For each solution, we compute static opcode counts from compiled bytecode and dynamic opcode counts from runtime traces on each test. These counts are transformed into normalized PMFs:
\[
  p_i\;=\;\frac{c_i}{\sum_{j} c_j} \quad ,
  \qquad
  q_i\;=\;\frac{w_i\,c_i}{\sum_{j} w_j c_j} \quad .
\]
This yields the following tensors: \textbf{Static PMFs:} $P \in \mathbb{R}^{m \times d}$ (structural), $Q \in \mathbb{R}^{m \times d}$ (cost-weighted); and \textbf{Dynamic PMFs:} $D \in \mathbb{R}^{r \times m \times d}$, $C \in \mathbb{R}^{r \times m \times d}$.

\subsubsection{Measuring Divergence via Statistical Distances}
\label{subsubsec:statistical_measures}

To compare PMFs, we employ two distinct yet complementary families of statistical metrics. This dual approach provides a comprehensive and robust assessment of stability by capturing divergence from two different perspectives and helps us triangulate our findings, ensuring that the observed stability patterns are a genuine characteristic of the model's behavior rather than an artifact of a specific metric's properties.

\begin{itemize}
    \item \textbf{Jensen--Shannon Divergence (JSD):} A symmetric, smoothed variant of KL-divergence, bounded in $[0,1]$~\cite{menendez1997jensen}. Our JSD-based metrics quantify stability through the lens of average pairwise dissimilarity.
    \item \textbf{Normalized Total Variance ($\tau$):} A metric derived from the trace of the covariance matrix of the PMF vectors. We normalize it using a theoretical upper bound, ensuring the result is constrained within $[0,1]$. This variance-based approach essentially evaluates how tightly the entire set of generated solutions is clustered around a common mean.
\end{itemize}

\subsubsection{Interpretability and Boundedness}

For SCTD and DCTD:

\begin{itemize}
    \item A score of \textbf{0} implies perfect stability (identical opcode distributions across all solutions).
    \item A score of \textbf{1} implies maximal instability (maximum divergence in opcode profiles).
\end{itemize}

We provide formal proofs of these bounds in Section~\ref{sec:metrics}.

\subsection{Metrics Definition}
\label{sec:metrics}


\subsubsection{Static Canonical Trace Divergence (SCTD)}
SCTD measures algorithmic divergence by comparing opcode distributions from compiled bytecode, quantifying the variety of generated solutions \emph{before} execution.

\paragraph{JSD-Based Formulation (SCTD\_JSD)}
This formulation computes the average pairwise JSD between the PMFs of all solutions. The final score is a convex combination of the divergence from the structural ($P$) and the cost-weighted PMFs ($Q$).
{\scriptsize
\begin{align*}
  \mathrm{SCTD\_JSD} = \; &\alpha\,\underbrace{\Bigl[\tfrac{2}{m(m-1)}\sum_{1\le s<t\le m} \mathrm{JSD}(p_s,p_t)\Bigr]}_{\substack{\text{\tiny from structural PMFs } (P)}} \\
  & + (1-\alpha)\,\underbrace{\Bigl[\tfrac{2}{m(m-1)}\sum_{1\le s<t\le m} \mathrm{JSD}(q_s,q_t)\Bigr]}_{\substack{\text{\tiny from cost-weighted PMFs } (Q)}}
\end{align*}
}
\paragraph{Covariance-Based Formulation (SCTD\_$\tau$)}
This formulation measures the dispersion of the PMFs around their mean. Let $X_P$ and $X_Q$ be random variables for a PMF chosen uniformly from the rows of tensors $P$ and $Q$, respectively. We define a normalized total variance score $\tau \in [0,1]$:
\[
  \boxed{\tau(X)\;=\;\frac{\tr\Sigma}{\;1-\|\mu\|_2^{2}\;}}, \quad \text{where } \mu=\E[X] \text{ and } \Sigma=\Cov(X).
\]
The normalization factor $1-\|\mu\|_2^2$ is the maximum possible trace of the covariance matrix for a random variable with mean $\mu$, as established in Theorem~\ref{thm:trace-bound}. The final score is a weighted combination of the $\tau$ scores:
\[
  \mathrm{SCTD\_\tau}\;=\;\alpha\,\tau(X_P)+(1-\alpha)\,\tau(X_Q).
\]

\subsubsection{Dynamic Canonical Trace Divergence (DCTD)}
DCTD extends the divergence metrics to \emph{run-time}.

\paragraph{JSD-Based Formulation (DCTD\_JSD)}
For each test case, we compute the average pairwise JSD between solutions. These per-test scores are then averaged across all tests and combined with the hyperparameter $\alpha$.
{\scriptsize
\begin{align*}
  \mathrm{DCTD\_JSD} = \; &\alpha\underbrace{\Bigl[\tfrac1r\sum_{j=1}^r \Bigl(\tfrac{2}{m(m-1)}\sum_{1\le s<t\le m}\mathrm{JSD}(p_{j,s}, p_{j,t})\Bigr)\Bigr]}_{\substack{\text{\tiny from structural PMFs } (P)}} \\
  & + (1-\alpha)\underbrace{\Bigl[\tfrac1r\sum_{j=1}^r \Bigl(\tfrac{2}{m(m-1)}\sum_{1\le s<t\le m}\mathrm{JSD}(q_{j,s}, q_{j,t})\Bigr)\Bigr]}_{\substack{\text{\tiny from cost-weighted PMFs } (Q)}}
\end{align*}
}
Here, $p_{j,s}$ and $q_{j,s}$ are the dynamic structural and cost-weighted PMFs for solution $s$ on test $j$.

\subsubsection{Covariance-Based Formulation (DCTD\_$\tau$)}
 Let $X_j$ and $Y_j$ be random variables representing a structural and cost-weighted PMFs chosen uniformly from the $m$ solutions for a given test case $j$.
{\scriptsize
\[
  \mathrm{DCTD\_\tau} = \alpha\underbrace{\Bigl[\tfrac1r\sum_{j=1}^r \tau(X_j)\Bigr]}_{\substack{\text{\tiny from structural PMFs } (P)}}
                     + (1-\alpha)\underbrace{\Bigl[\tfrac1r\sum_{j=1}^r \tau(Y_j)\Bigr]}_{\substack{\text{\tiny from cost-weighted PMFs } (Q}}.
\]
}
$\alpha \in [0,1]$ is an adjustable hyperparameter in all the metrics.
\subsubsection{Proofs of Boundedness}
A key property of our metrics is that they are bounded in $[0,1]$, making them easy to interpret.

\begin{theorem}[\textbf{JSD-Based Metrics are Bounded}]
\label{thm:composite-bounds}
Given that $\mathrm{JSD}(x,y) \in [0, 1]$ for any PMFs $x,y$, both SCTD\_JSD and DCTD\_JSD are bounded in $[0,1]$.
\end{theorem}
\begin{proof}
The SCTD\_JSD score is a convex combination of two terms, each being an average of JSD values. Since an average of values in $[0,1]$ is also in $[0,1]$, the entire expression is bounded in $[0,1]$. The DCTD\_JSD score involves an additional layer of averaging across test cases, which also preserves the $[0,1]$ bound. Therefore, both metrics are guaranteed to be in $[0,1]$.
\end{proof}

\begin{theorem}[\textbf{Sharp Upper Bound for $\tau$ Normalization}]\label{thm:trace-bound}
For a random PMF vector $X$ with mean $\mu=\E[X]$, the trace of its covariance matrix $\Sigma$ is bounded by:
\[
  \tr\Sigma\;\le\;1-\|\mu\|_2^{2}.
\]
\end{theorem}
\begin{proof}
The variance of any coordinate $X_i$ is $\Var(X_i)=\E[X_i^2]-\mu_i^2$. Since $0\le X_i\le1$, we have $X_i^2\le X_i$, which implies $\E[X_i^2] \le \E[X_i] = \mu_i$. Thus, $\Var(X_i) \le \mu_i-\mu_i^2$. Summing over all coordinates:
{\scriptsize
\[
  \tr\Sigma = \sum_i \Var(X_i) \;\le\;\sum_i (\mu_i-\mu_i^2)
              \;=\;\sum_i \mu_i - \sum_i \mu_i^2\;=\;1-\|\mu\|_2^2,
\]
}
where we use the fact that $\sum_i \mu_i = \sum_i \E[X_i] = \E[\sum_i X_i] = \E[1] = 1$. This result ensures that $\tau(X) \in [0,1]$. As SCTD\_$\tau$ and DCTD\_$\tau$ are constructed from convex combinations and averages of these normalized $\tau$ scores, they are also bounded in $[0,1]$.
\end{proof}
\begin{figure*}[htb]
\centering
\includegraphics[width=0.7\textwidth]{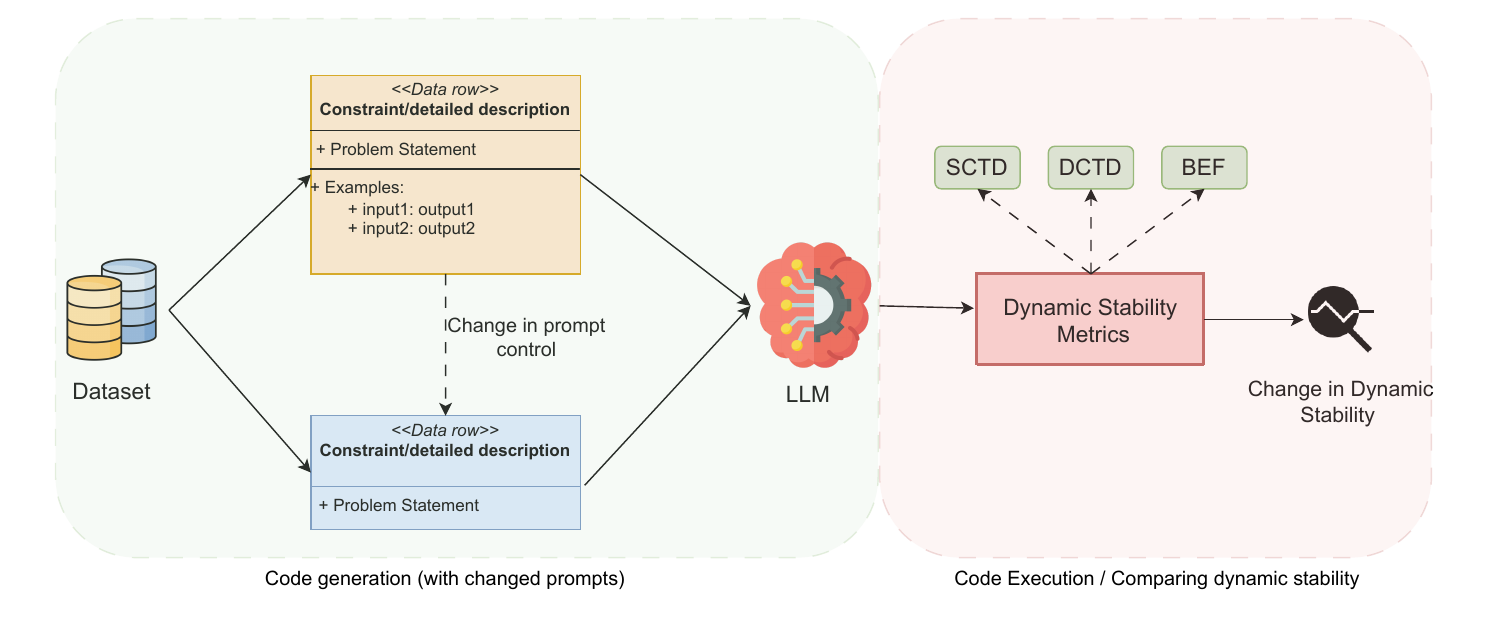}
\caption{Pipeline for evaluating the impact of prompt control on dynamic stability. Code is generated from a dataset using an LLM under varying prompt conditions, and the resulting outputs are assessed using dynamic stability metrics including DCTD, SCTD, and BEF.}
\label{fig:pipeline_prompt}
\vspace{-1em}
\end{figure*}
\subsubsection{Exploratory Analysis: Behavioral Expression Factor (BEF)}

While SCTD and DCTD provide direct measures of static and dynamic stability, their ratio offers a unique lens into the \emph{redundancy} of the generated code. We define the \textbf{BEF} as:
\begin{equation}
    \text{BEF} = \frac{\text{SCTD}}{\text{DCTD}}
\end{equation}
To avoid division-by-zero errors in cases of perfect dynamic stability (i.e., when DCTD = 0), we add a small constant $\epsilon = 10^{-9}$ to the denominator.
The BEF quantifies the extent to which static structural diversity among solutions leads to distinct runtime behavior.
The value of BEF reveals different patterns of alignment or mismatch between structure and behavior:

\begin{itemize}
    \item \textbf{BEF $\gg$ 1.} LLM generates many structurally different solutions (high SCTD) but low runtime variation (low DCTD). Many syntactically different solutions behave similarly, often due to unexercised or functionally equivalent code paths.

    \item \textbf{BEF $\approx$ 1.} Static and dynamic diversity are aligned. Distinct code structures lead to distinct behaviors, indicating meaningful variation and effective test coverage.

    \item \textbf{BEF $< 1$.} Runtime behavior diverges more than structure. Structurally similar solutions behave differently, often due to data-dependent branches triggered by test inputs.
\end{itemize}




\subsection{Research Questions}
\label{sec:rqs}

Our study is guided by the following research questions (RQs), each designed to probe a specific facet of the stability of LLM-generated code.

\begin{enumerate}[label=\textbf{RQ\arabic*.}, wide, labelindent=0pt]
    \item \textbf{How algorithmically stable is LLM-generated code across multiple completions?}
    With this RQ, our aim is to quantify the \textit{algorithmic stability} of the generated code. Does the model consistently generate similar algorithmic strategies, or does it produce structurally diverse but functionally equivalent solutions for the same prompt?

    \item \textbf{How dynamically stable is LLM-generated code at runtime across diverse inputs?}
    Here, we seek to measure the \textit{run-time stability} of generated solutions. Specifically, how consistent is the runtime behavior of generated solutions when executed on varying test inputs?

    \item \textbf{How effectively do SCTD, DCTD, and BEF capture the dimensions of code quality missed by traditional metrics?}
    This RQ assesses the validity of our proposed metrics.

    \item \textbf{How do prompts affect the algorithmic and dynamic stability?}
    Here, we investigate the potential for \textit{controlling} stability through prompt engineering. We explore whether prompt modifications systematically influence the stability of generated code.

    \item \textbf{How does temperature affect the algorithmic and dynamic stability of generated code?}
    
    Finally, this question examines the impact of a fundamental generation parameter. Specifically, what trade-offs arise between solution diversity and stability as temperature varies during generation. 
\end{enumerate}

\subsection{Experimental Setup}
\label{sec:experimental_setup}

Our experimental framework is built on code generation benchmarks and a controlled execution environment. This is essential to ensure that our measurements are both reliable and representative of the current capabilities of large language models.

\subsubsection{Datasets and Models}
\label{sec:datasets_models}

We evaluate on two benchmarks: CodeContests \cite{li2022competition} and BigO(Bench) \cite{chambon2025bigo}. BigO(Bench) is designed to study the link between algorithmic complexity and runtime behavior, making it well-suited for our analysis; we use its 311-problem test set. CodeContests provides a challenging functional correctness benchmark with difficulty ratings ('A' to 'D+'). We utilize the official validation and test splits. Both datasets include public and private unit tests, allowing us to identify functionally correct solutions using public tests and assess dynamic behavior using private ones. 
To ensure generalizability, we evaluate 11 LLMs spanning three functional categories (Table~\ref{tab:model_selection}): (1) \textbf{code generation models}, fine-tuned for synthesis; (2) \textbf{language/code models}, capable across modalities; and (3) \textbf{reasoning models}, optimized for multi-step problem solving. This diverse selection enables analysis across model scales, architectures, and specializations, covering both commercial and open-source models.

\begin{table}[h!]
  \centering
  \setlength{\aboverulesep}{0pt}
  \setlength{\belowrulesep}{0pt}
  \renewcommand{\arraystretch}{0.9}
  \setlength{\tabcolsep}{2pt}
  \scriptsize
  \caption{Large language models selected for our stability evaluation, grouped by source and specialization. Abbreviations are used in plots and figures.}
  \label{tab:model_selection}
  \begin{tabular}{@{} l l l l l @{}}
    \toprule
    \textbf{Model} & \textbf{Abbreviation} & \textbf{Params} & \textbf{Context} & \textbf{Developer} \\
    \midrule

    \multicolumn{5}{@{}>{\columncolor{gray!20}}l}{\textbf{Commercial Models}} \\
    \multicolumn{5}{@{}l}{\textbf{\textit{Language/Code Models}}} \\
    GPT-3.5-turbo-instruct     & GPT-3.5       & N/A   & 16k   & OpenAI \\
    GPT-4o                     & GPT-4o        & N/A   & 128k  & OpenAI \\
    Claude-3.7-Sonnet          & Claude-3.7-S  & N/A   & 200k  & Anthropic \\

    \multicolumn{5}{@{}l}{\textbf{\textit{Reasoning Model}}} \\
    GPT-o4-mini                & GPT-o4-m      & N/A   & 128k  & OpenAI \\

    \midrule
\multicolumn{5}{@{}>{\columncolor{gray!20}}l}{\textbf{Open-Source Models}} \\
    \multicolumn{5}{@{}l}{\textbf{\textit{Code Models}}} \\
    Qwen2.5-Coder-7B           & Qwen-7B-C     & 7B    & 64k   & Alibaba Cloud \\
    CodeLlama-7B-Instruct      & CodeLlama-7B-It & 7B  & 16k   & Meta \\
    Codestral-22B              & Codestral-22B & 22B   & 32k   & Mistral AI \\

    \multicolumn{5}{@{}l}{\textbf{\textit{Language/Code Models}}} \\
    Llama-3.1-8B               & Llama3.1-8B   & 8B    & 128k  & Meta \\
    Mistral-7B-v0.3            & Mistral-7B    & 7B    & 32k   & Mistral AI \\

    \multicolumn{5}{@{}l}{\textbf{\textit{Reasoning Models}}} \\
    DeepSeek-R1-Distill-Qwen-32B & DS-Qwen-32B & 32B   & 128k  & DeepSeek AI \\
    DeepSeek-R1-Distill-Llama-70B & DS-Llama-70B & 70B & 128k  & DeepSeek AI \\

    \bottomrule
  \end{tabular}
\end{table}

\subsubsection{Metric Calculation and Execution Protocol}
\label{sec:metric_calculation}

For each model–problem pair, we generate five solutions ($n = 5$) at three temperatures: 0.0, 0.7, and 0.95. This sampling strategy is intentionally designed to probe the model’s probabilistic output space, where a stable model is expected to produce solutions of similar complexity. To further analyze the effect of temperature in finer detail, we performed a systematic temperature scan for GPT-3.5-turbo-instruct from 0.0 to 2.0 in increments of 0.2. We set the private tests $r = 10$ (some problems have fewer than 10 private tests; for these, we use the maximum available), and $\alpha = 0.5$ for all our metrics, these variables are formally defined in Section~\ref{sec:metrics}.


\subsubsection{Execution and Measurement Protocol} 
To ensure both security and reproducibility, functionally correct solutions are executed in a secure, single-threaded sandbox using the private test suite, with a 20s timeout per test to handle non-termination. To ensure accurate opcode tracing via `sys.settrace`, we avoid multiprocessing.


\section{Results}




We report only \texttt{$all\_success$} results; additional cases are available on our anonymized GitHub\footnotemark[1]. SCTD and DCTD are presented as percentages of their theoretical maxima, derived from the dimensionless quantities in Section~\ref{sec:metrics}, scaled by 100 and denoted with \texttt{\_pct} or percentage symbols (\%).

\subsubsection{Measuring Static and Dynamic Divergence}
\begin{figure}[H] 
  \centering
  \includegraphics[width=\columnwidth]{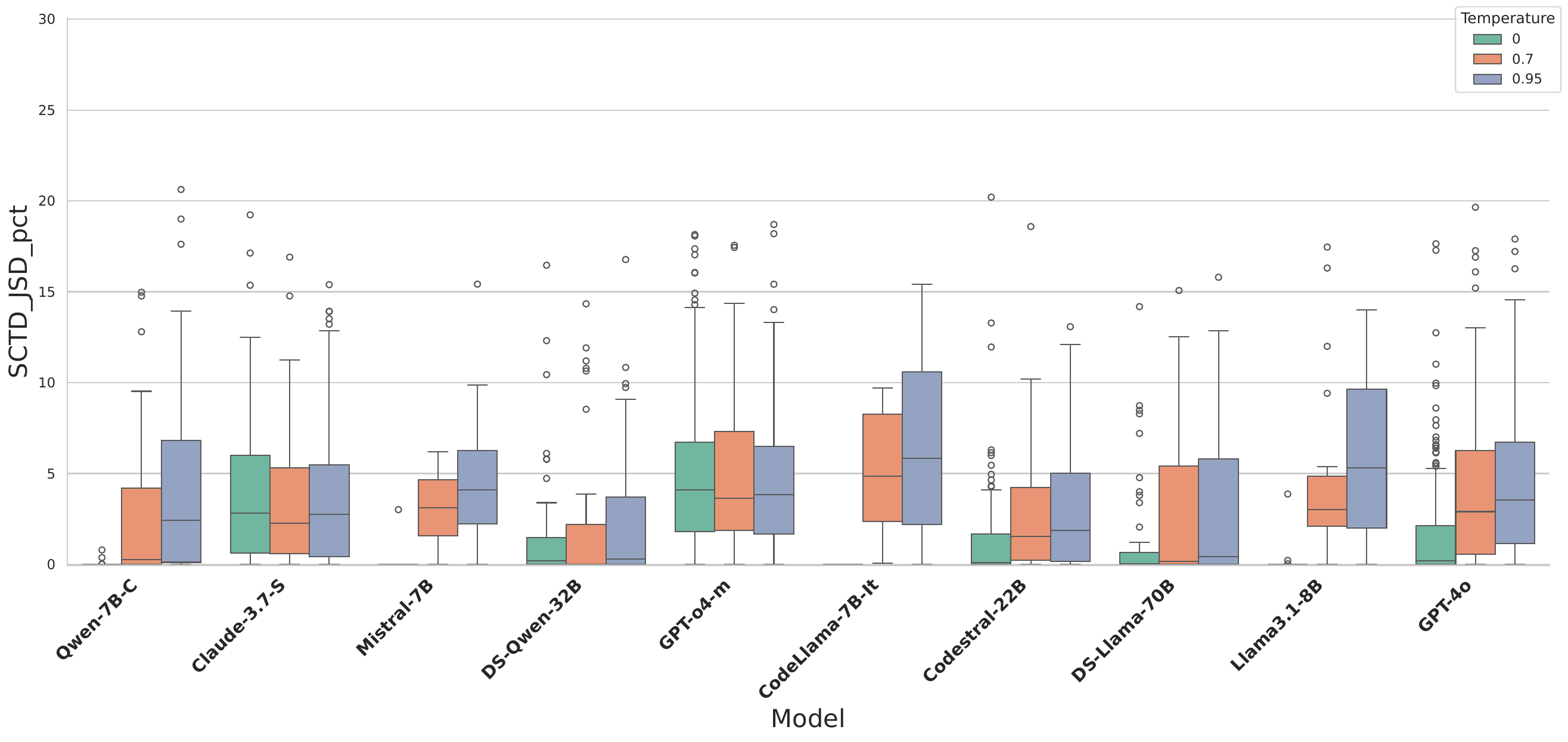}
  \caption{SCTD\_JSD for BigObench at different temperatures}
  \label{fig:sctd_jsd}
  \vspace{-2em}
\end{figure}
\begin{figure}[H] 
  \centering
  \includegraphics[width=\columnwidth]{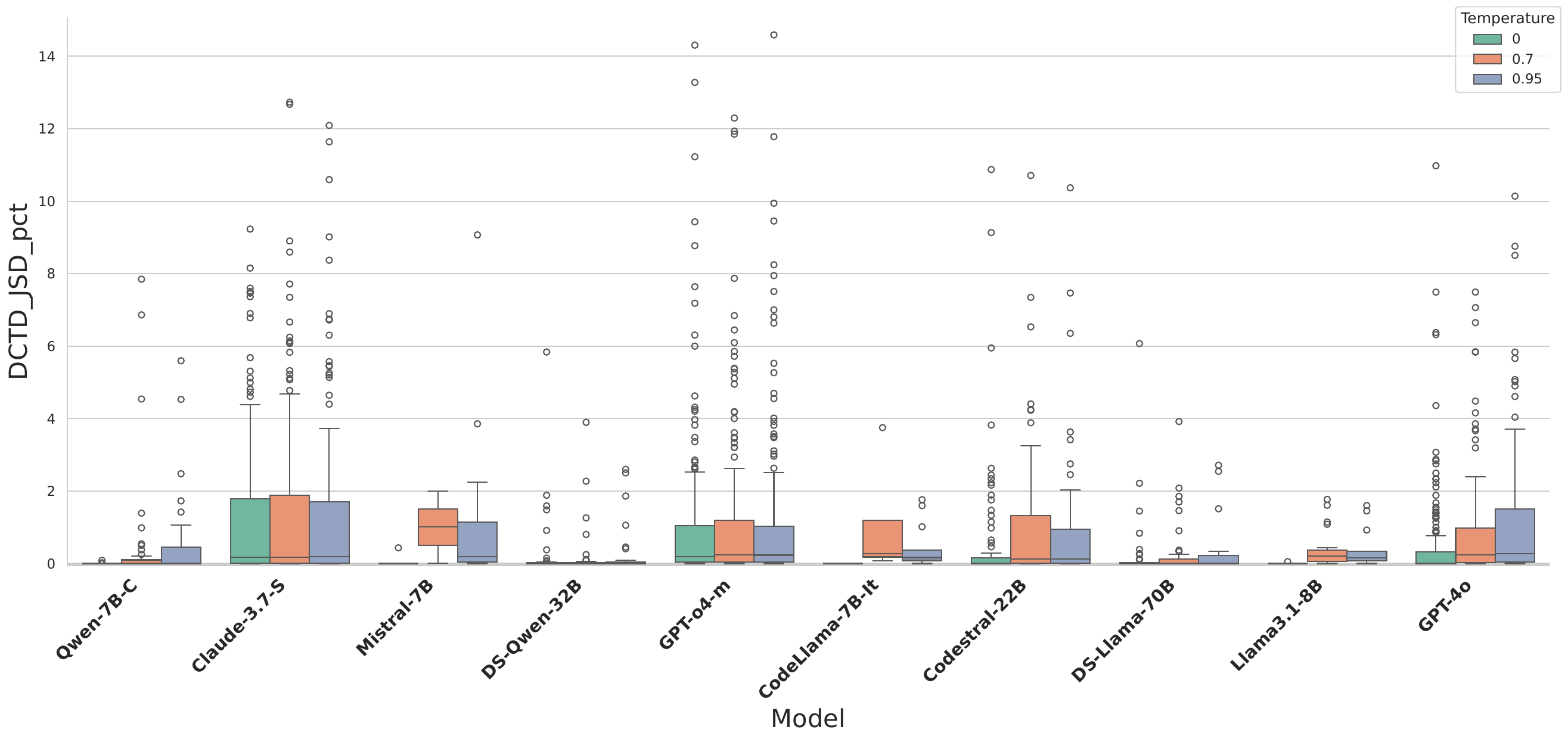}
  \caption{DCTD\_JSD for BigOBench at different temperatures}
  \label{fig:dctd_jsd}
  \vspace{-1em}
\end{figure}
Figures~\ref{fig:sctd_jsd} and \ref{fig:dctd_jsd} illustrate the static and dynamic divergence (for functionally correct solutions). As shown in Figure~\ref{fig:sctd_jsd}, the SCTD consistently falls in the range of 3--5\% of the maximum possible divergence for most models. This indicates that LLMs produce a variety of algorithmic structures even when generating correct code.
More critically, Figure~\ref{fig:dctd_jsd} shows that the DCTD is around 1\%  of the maximum. While these percentages may seem small, we highlight that they are the average metrics, and the theoretical upper bound represents a state of maximal chaos; moreover, as can be seen in the box plots, it is the anomalous points—the points that diverge too far (in ranges of approximately 20\% for SCTD and approximately 10--15\% for DCTD of the respective theoretical upper bounds) that are more concerning here. Even if some of the cases are problematic, they can create runtime bottlenecks. To provide a more granular view than the figures allow, Table~\ref{tab:main_results} presents a detailed snapshot of our results at temp=0.7. The strong agreement between our two metric types (JSD and $\tau$) validates the measurement approach. Detailed analysis further reveals that the average numbers and anomalous cases both increase drastically for for \texttt{some\_success} and \texttt{all\_failure} cases, which cannot be completely ignored, as \texttt{some\_success} is the most representative of a real-world situation where an LLM is run multiple times until it generates a working solution (more details are in the anonymized Github\footnotemark[1] due to space constraints).

\begin{figure*}[htbp]
    \centering
    \includegraphics[width=\linewidth]{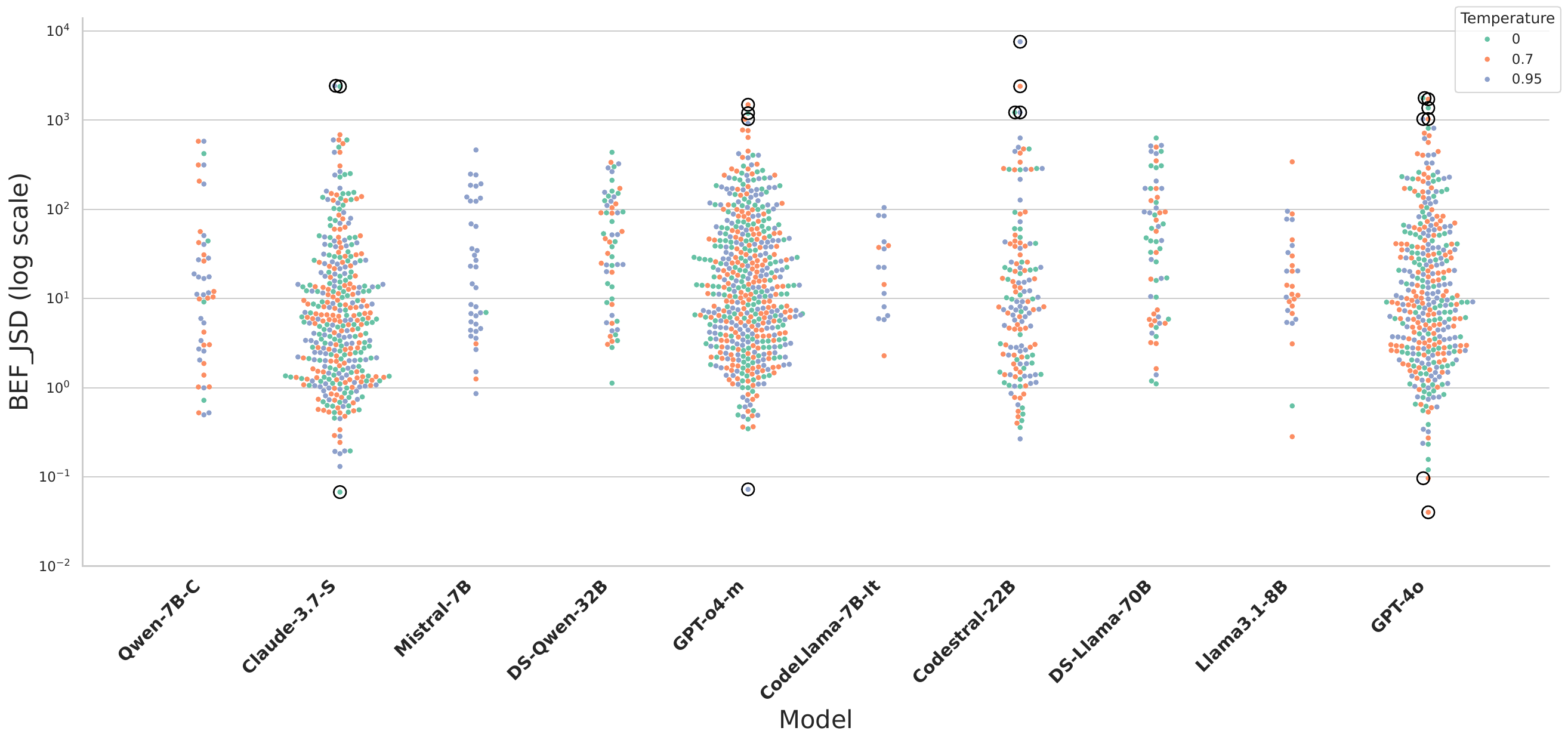}
    \caption{BEF\_JSD BigOBench for models at different temperatures}
    \label{fig:BEF_swarm}
    \vspace{-1em}
\end{figure*}
\begin{figure*}[htbp]
    \centering
    \includegraphics[width=0.8\linewidth]{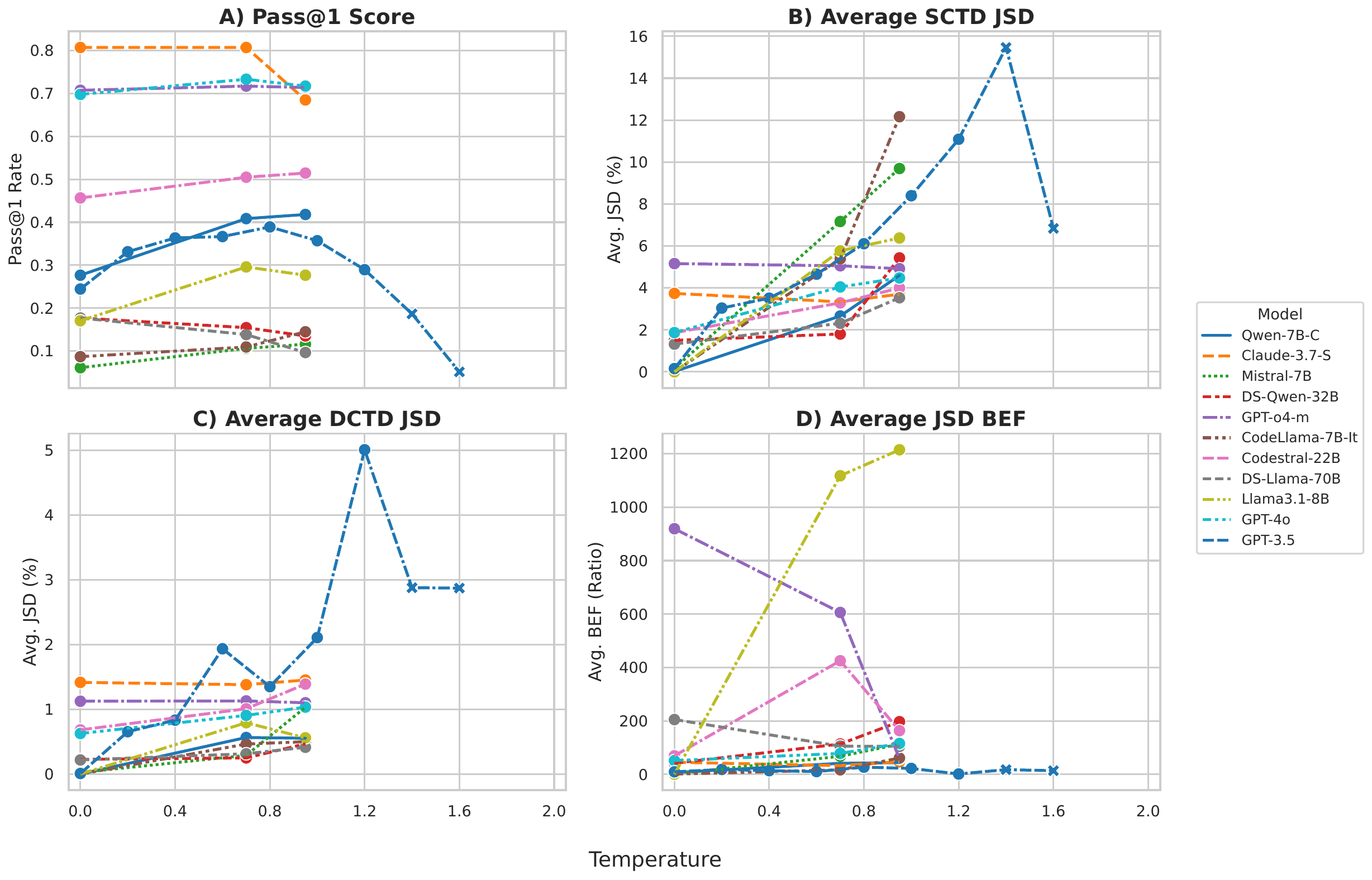}
    \caption{Change in pass@1 with temperature compared to change in dynamic stability}
    \label{fig:temp_stability}
    \vspace{-1em}
\end{figure*}

\begin{tcolorbox}[leftrule=0mm,rightrule=0mm,toprule=0mm,bottomrule=0mm,left=0pt,right=0pt,top=0pt,bottom=0pt,title={}]
\underline{\bf Answer to RQ1 \& RQ2:} LLM-generated code exhibits significant algorithmic (SCTD) and dynamic (DCTD) instability, even when all generated solutions are functionally correct. Our metrics successfully capture this divergence, with the Behavioral Expression Factor (BEF) revealing critical cases where structurally similar code behaves unpredictably at runtime, posing a direct risk to production systems.
\end{tcolorbox}

The Anomalous points represent variation in how these functionally equivalent programs execute differently.
\subsubsection{Interpreting Divergence with the Behavioral Expression Factor (BEF)}
The ratio of static to dynamic divergence, the BEF, provides a powerful lens for interpreting this instability. The distribution of BEF values in Figure~\ref{fig:BEF_swarm} highlights two particularly insightful scenarios:
\textbf{BEF > 1000}: This indicates SCTD >> DCTD, revealing a high degree of \textit{functional redundancy}. The model generates many structurally different solutions, but they converge to nearly identical runtime behaviors. This suggests the structural diversity is superficial and that the benchmark's tests may not be diverse enough to exercise unique code paths. \textbf{BEF < 0.1}: This is a critical indicator of risk where DCTD >> SCTD. The generated solutions are structurally very similar, yet they exhibit dramatically different runtime behaviors. A subtle change in the code triggers a cascade of divergent execution paths, creating unpredictable performance bottlenecks and vulnerabilities.

\subsubsection{Effectiveness and Validity of the Stability Metrics}
\label{subsubsec:rq_validity}

Our empirical results provide a clear and direct answer to the question of our metrics’ validity and effectiveness. The sensitivity of these metrics is illustrated in Figures~\ref{fig:sctd_jsd} and \ref{fig:dctd_jsd}. For any given problem, when evaluated across multiple runs, our metrics are capable of indicating dynamic instability in the generated solutions. This instability signals that \textbf{one or more of the solutions are algorithmically superior (SCTD) or exhibit better runtime behavior (DCTD)}, offering a practical cue for further optimization. While identifying the absolute best solution remains inherently uncertain, since there is no definitive way to determine the optimal time complexity for any randomly given problem. Our metrics provide a sufficient indication of potential failure points. This is especially valuable in the context of deploying LLMs for automated software engineering and code generation tasks, where such signals can guide refinement and reliability assessment. Finally, their diagnostic power is proven by the Behavioral Expression Factor's (BEF) capacity to pinpoint specific, high-risk failure modes(Figure~\ref{fig:BEF_swarm}). This body of evidence grounds the utility of our framework not in theory, but in its proven ability to uncover and quantify otherwise invisible properties of LLM-generated code.
\begin{tcolorbox}[leftrule=0mm,rightrule=0mm,toprule=0mm,bottomrule=0mm,left=0pt,right=0pt,top=0pt,bottom=0pt,title={}]
\underline{\bf Answer to RQ3:} Our metrics are effective because they provide an \textbf{actionable signal for optimization}. By quantifying stability with opcode traces, they reveal the existence of algorithmically superior choices within a set of correct solutions, pinpointing both runtime risks and optimization opportunities missed by traditional evaluations.
\end{tcolorbox}

\subsubsection{Independence from Traditional Code Metrics}
A foundational step is to ensure our proposed metrics capture a novel aspect of code quality not already covered by existing techniques. To this end, we conducted a Pearson correlation analysis against established, syntax-focused metrics like AST-similarity\cite{continuous-eval},  Tree Similarity of Edit Distance (TSED)\cite{song2024revisiting}, CodeBLEU\cite{ren2020codebleu}, and n-gram similarity\cite{continuous-eval}. For this comparison to be meaningful, a methodological adjustment is necessary: our metrics quantify \textit{divergence} (where 0 indicates perfect stability), while traditional similarity metrics operate on an inverted scale (where 1 indicates perfect similarity). Therefore, for this analysis only, we use the transformed stability scores \textbf{1-SCTD} and \textbf{1-DCTD} to align the interpretation across all metrics, where a higher value consistently signifies greater stability or similarity.
\begin{figure}[H] 
  \centering
  \includegraphics[width=\columnwidth]{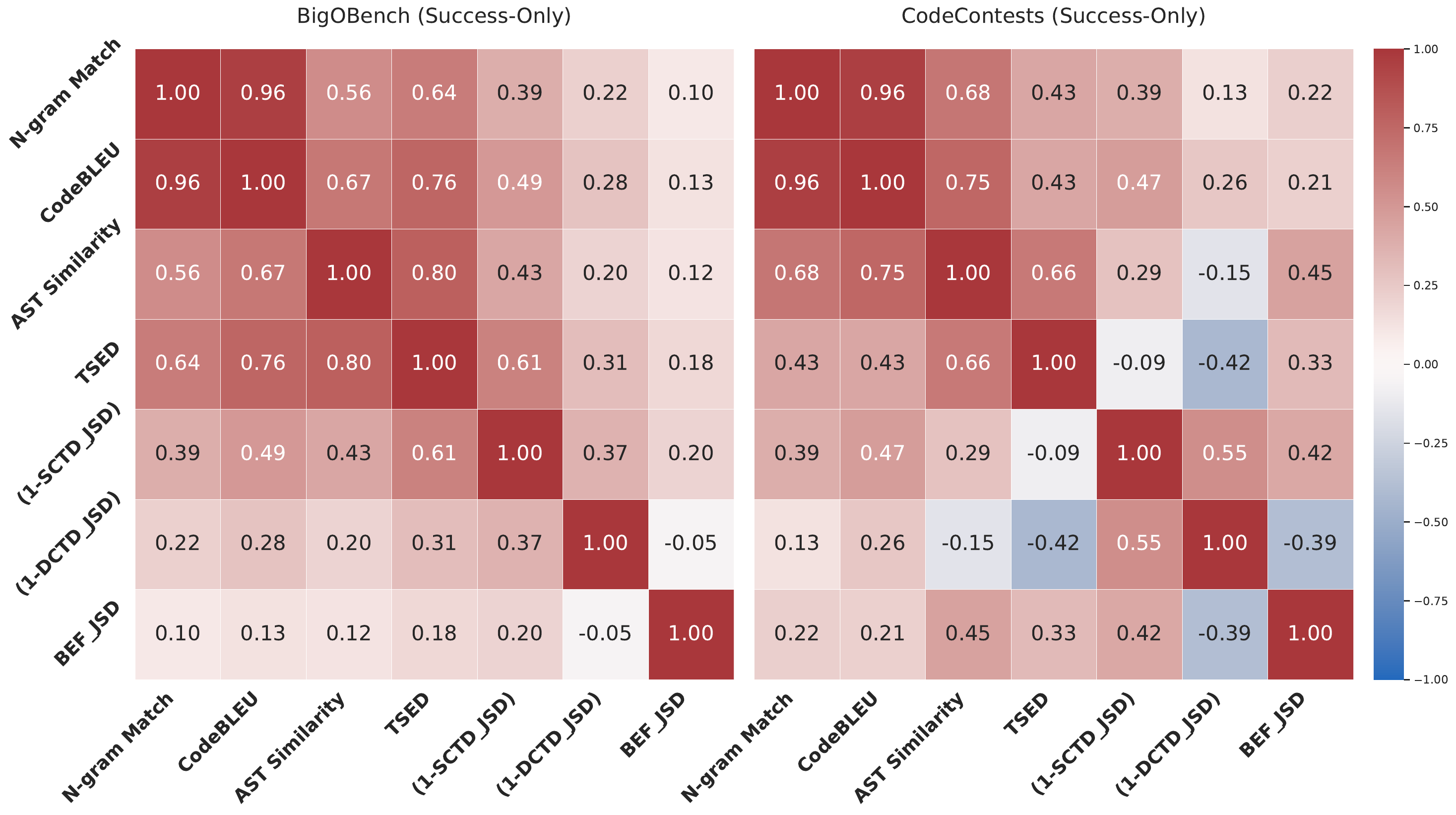}
  \caption{GPT-4o Pearson correlation heatmap comparing SCTD\_JSD, DCTD\_JSD and BEF\_JSD with other diversity metrics.}
  \label{fig:pearson_corr_jsd}
  \vspace{-1em}
\end{figure}
The resulting heatmap in Figure~\ref{fig:pearson_corr_jsd} shows consistently low correlation coefficients. This demonstrates that our framework quantifies crucial dimensions of algorithmic and runtime variance, which is largely orthogonal to static structural similarity. The same low-correlation pattern holds when analyzing our \texttt{$\tau$} based metrics (more details in the anonymized Github\footnotemark[1]).

\subsubsection{Can prompt engineering be a reliable lever for controlling stability}
To investigate this relationship, we analyzed the effect of removing in-context examples from prompts as a proxy for reducing prompt control, as shown in Figure~\ref{fig:pipeline_prompt}. While pass@1 scores generally decreased across most models following the removal, as detailed in Table~\ref{tab:change_analysis}, the impact on our stability metrics exhibited considerable variability. Specifically, we observed no consistent pattern: for certain models and datasets, stability improved as indicated by negative changes in SCTD and DCTD, whereas for others, it deteriorated significantly, reflected in positive changes in these metrics.
We acknowledge that prompt engineering is a broad and diverse field, and we do not assert that the observed correlations are universally applicable across all prompt engineering and control methods. However, our study demonstrates that the interdependence between prompt control and stability is non-trivial and warrants further examination. A comprehensive, in-depth study will be necessary to rigorously analyze the effect of prompt engineering on dynamic stability across diverse settings.
\begin{tcolorbox}[leftrule=0mm,rightrule=0mm,toprule=0mm,bottomrule=0mm,left=0pt,right=0pt,top=0pt,bottom=0pt,title={}]
\underline{\bf Answer to RQ4:} Simple prompt engineering, such as removing in-context examples, does not have a predictable effect on the stability of generated code. The impact is random across models, indicating that \textbf{behavioral stability is not an intrinsic learning objective} for current LLMs and cannot be reliably managed through basic prompt manipulation alone.
\end{tcolorbox}
This lack of a direct, predictable relationship strongly suggests that stability must be incorporated into the model training or fine-tuning process as a first-class objective, rather than being treated as a post-hoc property to be managed via prompting.
\subsubsection{The Influence of Temperature and the "Penalty of Instability"}
\label{subsubsec:rq_temperature}

Our investigation into the effect of the temperature sampling parameter reveals a fundamental tension between achieving functional correctness and maintaining behavioral stability.

\begin{tcolorbox}[leftrule=0mm,rightrule=0mm,toprule=0mm,bottomrule=0mm,left=0pt,right=0pt,top=0pt,bottom=0pt,title={}]
\underline{\bf Answer to RQ5:} Increasing the sampling temperature creates a direct trade-off: while it may improve the \texttt{pass@1} rate by exploring a wider solution space, it consistently degrades dynamic stability. We term this phenomenon the \textbf{"Penalty of instability,"} where the search for correctness inadvertently introduces significant and costly runtime unpredictability.
\end{tcolorbox}

As illustrated in Figure~\ref{fig:temp_stability}, our results reveal a consistent pattern across nearly all models: moderately increasing temperature improves the \texttt{pass@1} rate, but this gain is coupled with a steady rise in both algorithmic and dynamic instability (SCTD and DCTD). This highlights an inherent trade-off between correctness and predictability. To visualize this phenomenon, particularly at the extremes of stochasticity, our plots employ a specific methodological distinction. The primary stability metrics, denoted by solid circles (\textbullet), are calculated exclusively from the \texttt{all\_success} cohort where all generated solutions are functionally correct. However, at very high temperatures (e.g., T > 1.2 for \texttt{gpt-3.5}), models often fail to produce any \texttt{all\_success} sets due to increasingly incoherent outputs. In these regimes, we employ a fallback methodology, calculating metrics on the \texttt{some\_success} cohort and representing these data points with cross markers (X) to cover the full range of temperature.

\begin{table*}[t]
\centering
\caption{Comparative Analysis of LLMs across Functional, Structural, and Behavioral Stability Metrics}
\label{tab:main_results}
\footnotesize 
\setlength{\tabcolsep}{3pt} 
\begin{tabular}{@{}lcccccccccccccc@{}}
\toprule
& \multicolumn{2}{c}{\textbf{pass@1}}
& \multicolumn{2}{c}{\textbf{\begin{tabular}[c]{@{}c@{}}SCTD\_JSD \\ (\%)\end{tabular}}}
& \multicolumn{2}{c}{\textbf{\begin{tabular}[c]{@{}c@{}}SCTD\_TAU \\ (\%)\end{tabular}}}
& \multicolumn{2}{c}{\textbf{\begin{tabular}[c]{@{}c@{}}DCTD\_JSD \\ (\%)\end{tabular}}}
& \multicolumn{2}{c}{\textbf{\begin{tabular}[c]{@{}c@{}}DCTD\_TAU \\ (\%)\end{tabular}}}
& \multicolumn{2}{c}{\textbf{BEF\_JSD}}
& \multicolumn{2}{c}{\textbf{BEF\_TAU}} \\
\cmidrule(lr){2-3} \cmidrule(lr){4-5} \cmidrule(lr){6-7} \cmidrule(lr){8-9} \cmidrule(lr){10-11} \cmidrule(lr){12-13} \cmidrule(lr){14-15}
\textbf{Model} & \textbf{BigO} & \textbf{CC} & \textbf{BigO} & \textbf{CC} & \textbf{BigO} & \textbf{CC} & \textbf{BigO} & \textbf{CC} & \textbf{BigO} & \textbf{CC} & \textbf{BigO} & \textbf{CC} & \textbf{BigO} & \textbf{CC} \\
\midrule

\multicolumn{15}{@{}l}{\cellcolor{gray!20}\textbf{Commercial Models}} \\
\textbf{\textit{Language/Code Models}} \\
\texttt{GPT-4o} & 0.73 & 0.17 & 4.05 & 1.88 & 42.35 & 13.90 & 0.91 & 0.54 & 14.44 & 5.92 & 78.45 & 675.70 & 2.48e+6 & 2.47e+3 \\
\texttt{Claude-3.7-Sonnet} & \textbf{0.81} & \textbf{0.23} & 3.34 & 4.07 & 36.43 & 49.09 & 1.38 & 1.01 & 22.26 & 12.75 & 32.86 & \textbf{1.94e+3} & 2.78e+6 & \textbf{4.91e+6} \\
\texttt{gpt-3.5-turbo-instruct} & 0.38 & - & 5.38 & - & 56.64 & - & \textbf{1.64} & - & \textbf{31.37} & - & 18.81 & - & 20.44 & - \\
\addlinespace
\textbf{\textit{Reasoning Model}} \\
\texttt{GPT-o4-mini} & 0.72 & 0.11 & 5.05 & 2.44 & 50.10 & 23.12 & 1.13 & 0.48 & 18.56 & 9.14 & 605.96 & 38.22 & 3228.66 & 32.14 \\
\midrule

\multicolumn{15}{@{}l}{\cellcolor{gray!20}\textbf{Open Source Models}} \\
\textbf{\textit{Code Models}} \\
\texttt{Qwen2.5-Coder-7B} & 0.41 & 0.05 & 2.66 & 1.20 & 25.88 & 7.48 & 0.57 & 0.00 & 6.93 & 0.00 & 41.41 & 693.21 & \textbf{2.57e+7} & 1.84e+3 \\
\texttt{CodeLlama-7B-Instruct} & 0.11 & 0.01 & 5.39 & 7.87 & 56.08 & -- & 0.46 & 0.55 & 5.56 & -- & 17.13 & 31.04 & 22.06 & -- \\
\texttt{Codestral-22B} & 0.50 & 0.10 & 3.29 & 2.34 & 44.66 & 17.05 & 1.01 & 1.94 & 16.08 & \textbf{19.69} & 425.28 & 26.69 & 1.40e+6 & 29.15 \\
\addlinespace
\textbf{\textit{Language/Code Models}} \\
\texttt{Llama-3.1-8B} & 0.30 & 0.06 & 5.77 & \textbf{12.02} & \textbf{84.67} & \textbf{185.91} & 0.79 & 1.05 & 9.49 & 13.43 & \textbf{1117.58} & 11.46 & 2206.19 & 13.85 \\
\texttt{Mistral-7B-v0.3} & 0.11 & 0.01 & \textbf{7.17} & 7.58 & 65.80 & -- & 0.29 & \textbf{22.25} & 4.82 & -- & 67.80 & 0.36 & 303.99 & -- \\
\addlinespace
\textbf{\textit{Reasoning Models}} \\
\texttt{DeepSeek-R1-Distill-Llama-70B} & 0.14 & 0.01 & 2.31 & 1.25 & 19.19 & -- & 0.31 & 0.03 & 2.85 & -- & 105.80 & 18.62 & 116.20 & -- \\
\texttt{DeepSeek-R1-Distill-Qwen-32B} & 0.15 & 0.01 & 1.80 & 10.25 & 21.68 & -- & 0.25 & 8.42 & 3.06 & -- & 113.72 & 1.11 & 279.28 & -- \\

\bottomrule
\multicolumn{15}{@{}p{\dimexpr\textwidth-2\tabcolsep}@{}}{\footnotesize \textbf{Metrics Guide}: Comparing stability metrics with functional correctness (\textbf{pass@1}) in BigOBench (BigO) and Coding Contests (CC). All results at temperature 0.7. ‘--’ indicates the model failed to generate at least two valid solutions for any dataset problem, while ‘-’ indicates the model was not run.}
\end{tabular}
\end{table*}

\begin{table*}[t]
\centering
\caption{Analysis of Metric Changes upon Removal of In-Context Examples. Values represent the absolute change (\textit{With examples to without examples}) for pass@1, SCTD/DCTD metrics and percentage change for BEF metrics.}
\label{tab:change_analysis}
\footnotesize 
\setlength{\tabcolsep}{3pt} 
\begin{tabular}{@{}lcccccccccccccc@{}}
\toprule
& \multicolumn{2}{c}{\textbf{\begin{tabular}[c]{@{}c@{}}$\Delta$ pass@1\end{tabular}}}
& \multicolumn{2}{c}{\textbf{\begin{tabular}[c]{@{}c@{}}$\Delta$ SCTD\_JSD\end{tabular}}}
& \multicolumn{2}{c}{\textbf{\begin{tabular}[c]{@{}c@{}}$\Delta$ SCTD\_TAU\end{tabular}}}
& \multicolumn{2}{c}{\textbf{\begin{tabular}[c]{@{}c@{}}$\Delta$ DCTD\_JSD\end{tabular}}}
& \multicolumn{2}{c}{\textbf{\begin{tabular}[c]{@{}c@{}}$\Delta$ DCTD\_TAU\end{tabular}}}
& \multicolumn{2}{c}{\textbf{\begin{tabular}[c]{@{}c@{}}$\Delta$ BEF\_JSD (\%)\end{tabular}}}
& \multicolumn{2}{c}{\textbf{\begin{tabular}[c]{@{}c@{}}$\Delta$ BEF\_TAU (\%)\end{tabular}}} \\
\cmidrule(lr){2-3} \cmidrule(lr){4-5} \cmidrule(lr){6-7} \cmidrule(lr){8-9} \cmidrule(lr){10-11} \cmidrule(lr){12-13} \cmidrule(lr){14-15}
\textbf{Model} & \textbf{BigO} & \textbf{CC} & \textbf{BigO} & \textbf{CC} & \textbf{BigO} & \textbf{CC} & \textbf{BigO} & \textbf{CC} & \textbf{BigO} & \textbf{CC} & \textbf{BigO} & \textbf{CC} & \textbf{BigO} & \textbf{CC} \\
\midrule

\multicolumn{15}{@{}l}{\cellcolor{gray!20}\textbf{Commercial Models}} \\
\textbf{\textit{Language/Code Models}} \\
\texttt{GPT-4o} & -0.11 & -0.01 & \negchange{0.48} & \poschange{0.02} & \poschange{14.05} & \negchange{13.79} & \poschange{0.57} & \poschange{0.06} & \negchange{5.13} & \negchange{5.85} & \poschange{46.16} & \negchange{98.16} & \poschange{99.81} & \negchange{99.59} \\
\texttt{Claude-3.7-Sonnet} & -0.18 & +0.01 & \poschange{0.20} & \negchange{0.39} & \negchange{47.19} & \negchange{48.74} & \negchange{0.10} & \poschange{0.57} & \poschange{13.01} & \negchange{12.53} & \poschange{57.00} & \negchange{98.72} & \poschange{900.00} & \negchange{99.95} \\
\texttt{gpt-3.5-turbo-instruct} & - & - & - & - & - & - & - & - & - & - & - & - & - & - \\
\addlinespace
\textbf{\textit{Reasoning Model}} \\
\texttt{GPT-o4-mini} & -0.12 & -0.01 & \poschange{0.44} & \poschange{0.59} & \negchange{21.93} & \negchange{22.70} & \poschange{0.51} & \poschange{0.60} & \poschange{9.15} & \negchange{8.97} & \negchange{95.28} & \poschange{21.58} & \negchange{65.22} & \poschange{63.01} \\
\midrule

\multicolumn{15}{@{}l}{\cellcolor{gray!20}\textbf{Open Source Models}} \\
\textbf{\textit{Code Models}} \\
\texttt{Qwen2.5-Coder-7B} & -0.14 & 0.00 & \poschange{1.66} & \negchange{1.18} & \poschange{7.89} & \negchange{7.48} & \poschange{0.91} & \poschange{0.03} & \negchange{0.01} & \poschange{0.01} & \poschange{2.72e+3} & \negchange{99.96} & \negchange{90.01} & \negchange{99.96} \\
\texttt{CodeLlama-7B-Instruct} & +0.01 & 0.00 & \poschange{1.70} & -- & \negchange{1.81} & -- & \negchange{0.15} & -- & \poschange{0.20} & xx & \poschange{9.31e+4} & -- & \negchange{9.05e+4} & -- \\
\texttt{Codestral-22B} & -0.10 & +0.01 & \poschange{0.01} & \negchange{0.62} & \poschange{17.25} & \negchange{16.93} & \poschange{0.54} & \negchange{0.11} & \negchange{18.99} & \negchange{19.48} & \negchange{98.03} & \poschange{1.48e+6} & \negchange{1.12e+6} & \poschange{1.08e+6} \\
\addlinespace
\textbf{\textit{Language/Code Models}} \\
\texttt{Llama-3.1-8B} & -0.07 & 0.00 & \poschange{1.63} & -- & \negchange{1.58} & -- & \poschange{0.20} & -- & \poschange{0.24} & -- & \negchange{97.51} & -- & \poschange{96.14} & -- \\
\texttt{Mistral-7B-v0.3} & -0.03 & -0.01 & \poschange{1.45} & -- & \poschange{1.39} & -- & \poschange{3.33} & -- & \negchange{3.50} & -- & \negchange{90.29} & -- & \negchange{91.33} & -- \\
\addlinespace
\textbf{\textit{Reasoning Models}} \\
\texttt{DeepSeek-R1-Distill-Llama-70B} & -0.01 & 0.00 & \poschange{0.34} & -- & \negchange{0.31} & -- & \negchange{0.18} & -- & \poschange{0.21} & -- & \negchange{12.43} & -- & \poschange{13.19} & -- \\
\texttt{DeepSeek-R1-Distill-Qwen-32B} & 0.00 & 0.00 & \poschange{0.78} & -- & \negchange{0.85} & -- & \poschange{0.12} & -- & \negchange{0.11} & -- & \negchange{54.13} & -- & \poschange{55.02} & -- \\

\bottomrule
\multicolumn{15}{@{}p{\dimexpr\textwidth-2\tabcolsep}@{}}{\footnotesize \textbf{Guide}: The table shows changes in dynamic stability after removing in-context examples. A \poschange{positive change (red)} indicates an increase in instability (higher variability), while a \negchange{negative change (green)} indicates an improvement in stability. All results are at temperature = 0.7.}
\end{tabular}
\end{table*}

\vspace{-1.5ex}
\section{Discussion}

\subsection{Sub-Symbolic Learning vs Formal Analysis}

For decades, a chasm has existed between symbolic AI, which operates on formal rules and can reason about concepts like Big-O notation explicitly, and sub-symbolic systems like LLMs, which learn from statistical patterns in data~\cite{marcus2020next, d2020neurosymbolic}. Our work doesn't just propose a new metric; it offers a practical bridge between these two worlds.

\begin{itemize}[leftmargin=*, nosep]
    \item \textbf{Implication.} For Python and by using opcode traces: a deterministic, low-level representation of a program's logic, our work provides a ``ground truth'' signal that is much closer to the formal semantics of a program than raw source code tokens. Training a model to minimize \textbf{SCTD/DCTD} is effectively teaching it to recognize and prefer computational patterns that are efficient, without explicitly teaching it algorithm analysis. It is an attempt to have the model \emph{induce} a theory of computational cost from empirical evidence.
    \item \textbf{Connecting the Dots.} This is a departure from classic prompt-based approaches like ``Chain-of-Thought'' which try to coax symbolic reasoning out of the model via text\cite{chambon2025bigo, wei2022chain}. Like other work that uses learning from the execution environment\cite{ouyang2022training, mankowitz2023faster}, we suggest a path to bake this reasoning into the model's weights through a feedback signal from the execution environment itself. It is less about making the model \emph{talk} like a computer scientist and more about making it \emph{behave} like one.
\end{itemize}

\subsection{"Algorithmic Debt" as a Quantifiable Risk in AI Agents}
The software engineering concept of "technical debt"\cite{cunningham1992wycash} describes how easy, short-term solutions lead to long-term maintenance costs. Our work provides the vocabulary and measurement tools for a more insidious, AI-specific variant: \emph{Algorithmic Debt}.
\begin{itemize}[leftmargin=*, nosep]
    \item \textbf{Implication.} As we move from single-function generation to autonomous AI agents that build and modify entire codebases (e.g., Devin\cite{devinai}, AutoGPT\cite{autogpt}), this problem will compound catastrophically. An agent might generate a slightly inefficient function (a small deposit of algorithmic debt). Then, it might build another function that calls that inefficient one in a loop, exponentially increasing the debt. The instability metrics SCTD and DCTD provide a way to quantify this debt at the point of creation.
    \item \textbf{Why this is Significant.} This reframes stability not as a "code quality" issue, but as a fundamental governor on the viability of long-running autonomous software engineering agents. An agent without the ability to manage its own algorithmic debt is doomed to build systems that will inevitably collapse under their own inefficiency. This work provides the sensor for an agent's "self-refactoring" drive\cite{le2019automated}.
\end{itemize}

\subsection{A New Perspective on the Geometry of the Solution Space}

Our finding of a "penalty of instability" with temperature offers a new insight into the geometry of an LLM's solution manifold. It suggests indeed that in the high-dimensional latent space of possible programs, the regions corresponding to "functionally correct" solutions are not smooth, well-behaved plateaus. Instead, they are potentially treacherous landscapes. This echoes findings from studies on neural network loss landscapes, where the geometry of minima (sharp vs. flat) correlates with generalization performance~\cite{draxler2018essentially, foret2020sharpness}

\begin{itemize}[leftmargin=*, nosep]
    \item \textbf{Implication.} Correct solutions with good performance might be located on narrow "ridges," surrounded by "cliffs" of high complexity or incorrect solutions. Increasing temperature allows the model to "jump" across valleys to find these ridges, but also increases the chance of falling off a cliff. The core challenge is not just finding these good solutions, but reshaping the manifold itself during training.
    \item \textbf{Connecting the Dots.} This connects to advanced research in representation learning. Using techniques like contrastive learning, we could use the stability metrics to pull representations of stable, correct solutions together in the latent space, while pushing unstable solutions far away~\cite{chen2020simple}. The goal would be to create a model where the most probable outputs are not just correct, but are drawn from a wide, smooth plateau of high-performance stability.
\end{itemize}



\section{Threats to Validity}

We acknowledge several limitations that may influence the interpretation and generalizability of our findings. \\
\textbf{Language Scope.} All experiments were conducted on Python, whose bytecode and runtime behavior differ from compiled or VM-based languages like C++, Rust, or Java. As such, our stability metrics may not generalize across languages without further validation. \\
\textbf{Test Suite Dependence.} DCTD and BEF rely on the coverage and diversity of test suites. While we highlight anomalies, the benchmarks lack asymptotic cases, limiting deeper behavioral analysis. \\
\textbf{Sample Size and Prompting.} Stability metrics are based on five generations per configuration, which may not fully capture output variance. Additionally, our prompt engineering analysis is limited to a basic strategy; more advanced techniques warrant future study.

\section{Related Work}
The evaluation of code-generating LLMs has progressed from simple functional correctness to sophisticated analyses of non-functional properties. We have broken down this evolution into three phases.\\
\textbf{Phase 1:} \textit{The Foundation of Functional Correctness.} The field's initial efforts were rightly centered on establishing functional viability. The dominant paradigm was, and to a large extent, still is functional correctness, quantified by the \texttt{pass@k} metric \cite{chen2021evaluating, li2022competition} on benchmarks like HumanEval \cite{chen2021evaluating} and MBPP \cite{austin2021program}. This phase proved that LLMs could generate code that produces correct outputs, but it treated all passing solutions as monolithic successes. \\ \textbf{Phase 2:} \textit{Deeper Quality Issues.} Recognizing that functionally identical solutions can differ greatly in quality, the research community began developing more nuanced metrics. An early and influential step was CodeBLEU \cite{ren2020codebleu}, which moved beyond input-output testing to incorporate syntactic similarity. This was followed by more advanced methods for measuring structural diversity. Techniques based on Abstract Syntax Tree (AST) similarity \cite{mastropaolo2021studying} and tree-based edit distances like TSED \cite{song2024revisiting} provided formal ways to quantify how different two pieces of code are in their fundamental structure. Parallel to this, empirical studies \cite{tian2023chatgpt} qualitatively confirmed that LLM solutions, while often correct, could be suboptimal or contain hidden flaws, further motivating the need for deeper analysis. \\
\textbf{Phase 3:} \textit{Learning from execution.} The most advanced approaches leverage dynamic signals from the execution environment itself. Prominent examples include AlphaDev \cite{mankowitz2023faster}, which used measured latency as a direct reward signal to discover faster algorithms, AlphaCode \cite{li2022competition}, which used pass/fail status on example tests to filter a massive set of candidates, and Green-Code \cite{ilager2025green}, which optimises for energy efficiency in code generation. More recently, the work of Chambon et al. \cite{chambon2025bigo} demonstrated that a model's prediction of algorithmic complexity correlates with execution time and that such complexity can be induced via prompting. However, this approach relies on a known ground-truth complexity for a given problem, which is often not available in real-world scenarios. Furthermore, it depends on specialized prompting techniques that are not standard practice for most users or autonomous agent systems, and does not study the run-time behavior of code. This reveals a critical gap that our work addresses.

\section{Conclusion}
This work establishes that functional correctness, while essential, represents an incomplete foundation for evaluating LLM-generated code in production environments. Our analysis reveals that models achieving identical correctness scores can exhibit dramatically divergent performance characteristics, creating substantial operational risks that current evaluation frameworks fail to capture. The phenomenon we term the "penalty of instability" demonstrates how the pursuit of functional accuracy can inadvertently generate solutions with unpredictable and costly runtime behavior.

These results call for a fundamental expansion of our evaluation paradigms beyond binary correctness metrics. We argue for elevating dynamic stability from a secondary consideration to a primary design principle in the development, evaluation, and deployment of AI-based code generation systems.



\balance
\bibliographystyle{ACM-Reference-Format}
\bibliography{main}

\appendix

\end{document}